\documentclass[aps,prl, twocolumn,groupedaddress,showpacs]{revtex4}

\usepackage{amssymb,amsfonts,amsmath}
\usepackage{graphicx}
\usepackage{amssymb}
\usepackage{epstopdf}
\usepackage{amssymb}
\usepackage{color}
\usepackage{multirow}

\newcommand{\comment}[1]{}

\newcommand{\micron}{\mu\text{m}}
\newcommand{\kt}{k_{\text{B}}T}

\newcommand{\DDP}{D_{\rm DP}}

\begin{document}

\title{Colloidal motility and localization under rectified diffusiophoresis}

\title{Colloidal motility and pattern formation under rectified diffusiophoresis}

\author{J\'{e}r\'{e}mie Palacci $^\dag$,
Benjamin Ab\'ecassis $^\ddag$,
C\'{e}cile Cottin-Bizonne $^\dag$,
Christophe Ybert $^\dag$
Lyd\'{e}ric Bocquet $^\dag$}
\email{lyderic.bocquet@univ-lyon1.fr}
\address{$^\dag$ LPMCN;  Universit\'e de Lyon; Universit\'e Lyon 1 and
CNRS, UMR 5586; F-69622 Villeurbanne, France \\
$^\ddag$ CEA, LETI, MINATEC, F38054 Grenoble, France}

\begin{abstract} 
{In this letter, we characterize experimentally the diffusiophoretic motion of colloids and $\lambda$-DNA toward higher concentration of solutes, 
using
microfluidic technology to build spatially- and temporally-controlled concentration gradients. We then demonstrate that segregation and spatial patterning of the particles can be achieved from temporal variations of the solute concentration profile.
This segregation takes the form of a strong trapping potential, stemming from an osmotically induced rectification mechanism of the solute time-dependent variations. Depending on the spatial and temporal symmetry of the solute signal, localization patterns with various shapes can be achieved. 
These results highlight the role of solute contrasts in out-of-equilibrium processes occuring in soft matter.
}
\end{abstract}
\maketitle

Diffusiophoresis is the mechanism by which particles and molecules drift along solute concentration gradients \cite{anderson,abecassis2008,Prieve87}.
It belongs to the general class of phoretic transport phenomena, such as electro- and thermo- phoresis \cite{Prost2009,Duhr06,Piazza08,Jiang09,Wurger09},
{\it i.e.} the migration of particles under gradients of an external thermodynamic variable (electric potential, temperature, ...).
Physically, the  diffusio-phoretic 
migration under solute gradients takes its origin in an osmotic pressure gradient occurring {\it within} the diffuse interface at the surface of the particle  \cite{anderson,AjdariBocquet2006,abecassis2008}: this leads to a drift velocity proportional to the solute gradient $ \nabla c$, in the form 
$V_{DP}=\mu \nabla c$, with $\mu$ a mobility. 
The  diffusiophoretic transport 
was recognized since the pioneering work of Derjaguin and later by Anderson and Prieve \cite{anderson}. However its implication in out-of-equilibrium phenomena occuring in soft matter systems -- in which solute contrasts are ubiquituous -- has been barely explored up to now.
Several works recently revealed the role played by transport induced by solute contrasts in out-of-equilibrium processes. 
It was {\it e.g.} shown to be at the origin of the  
strongly enhanced effective diffusion of colloids under solute gradients \cite{abecassis2008}
with potential implications in microfluidics \cite{abecassis2009}. Its possible interplay with other transport phenomena was also demonstrated, concuring for example to induce 
a sign reversal of thermophoretic transport of colloids in the presence of polymers \cite{Jiang09}. 
Its role in skin formation during evaporation of mixtures was suggested \cite{Kabalnov09}.
Finally, in the context of 
 the efforts to develop autonomous microsystems,
diffusiophoresis was used to harness chemical power to produce self-propelled artificial swimmers \cite{Golestanian},
using chemical gradients as driving forces.

\par
\begin{figure}[tb]
\includegraphics[width=8cm]{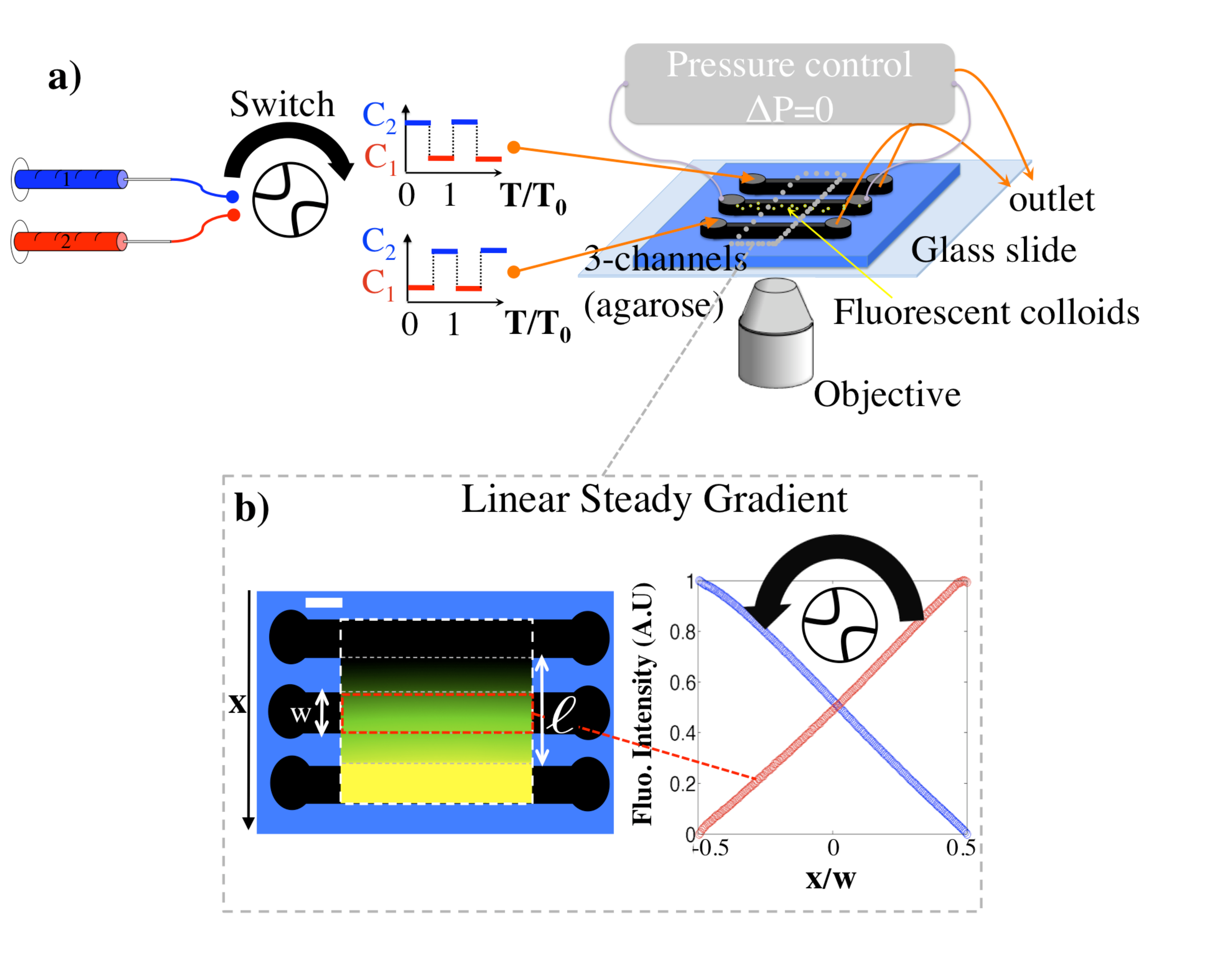}
\caption{{\it (a)} Experimental setup {\it (b)-(Left)} A 3-channel and gel matrix setup, superimposed on fluorescence intensity image measured for the stationary gradient 
of fluorescein with $C_1-C_2=10^{-4}$M  
($w=300\micron$, $\ell=800\,\mu$m, scale bar 300$\micron$).
%
{\it (b)-(Right)} Fluorescence intensity profiles measured in stationary state, showing the expected linear profile. 
Actuation of the microfluidic switch allows to invert the concentration gradient. 
\label{figure-1}
}
\end{figure}

In this paper we pursue the exploration of this phenomenon and demonstrate segregation and pattern formation of colloids and macromolecules on the basis of diffusiophoretic transport.
We evidence a localization process of particles --here, colloids and $\lambda$-DNA--, taking its origin in the rectification of the underlying time-dependent concentration variations of a solute specie.
The shape of the pattern is shown to be strongly dependent on the temporal symmetry of the underlying solute concentration signal.
A theoretical model based on a Smoluchowski description allows to reproduce the experimental observations.

%
%

The thorough exploration of diffusiophoresis requires the building-up of controlled salt concentration gradients. To this end, 
we have developed a microfluidic experimental setup sketched in Fig.\ref{figure-1},  inspired from Ref. \cite{Wu06}.
This set-up 
allows to impose stationary, as well as temporally switchable, solute gradients. 
A three channels device is molded in agarose gel, allowing free diffusion of solute without convective flow.
A double syringe pump with two solutions of salt of concentration $C_1$ and $C_2$  fills the two side channels. 
A microfluidic switch (Upchurch) allows to exchange the solute solutions in the side channels, leading to a time-dependent tuning of the gradient. Particles under investigation (colloids or $\lambda$-DNA) 
are located in the center channel. 
We use polystyren-carboxylate fluorescent colloids (F8888 200nm, Molecular Probes) in a Tris buffer (1mM, pH9), and
DNA  (48 kbp, Fermentas, Germany) 
 labelled with YOYO-1 (Invitrogen), in Tris-EDTA buffer (1mM, pH7.6). 
As a solute, we have considered here either fluorescein or bare salts --LiCl, NaCl and KCl--.
In a stationary regime, a linear profile of the solute concentration is achieved, 
as illustrated in Fig.~\ref{figure-1}-(b), 
with fluorescein as a benchmark. 
\begin{figure}[tbp]
\includegraphics[width=8cm]{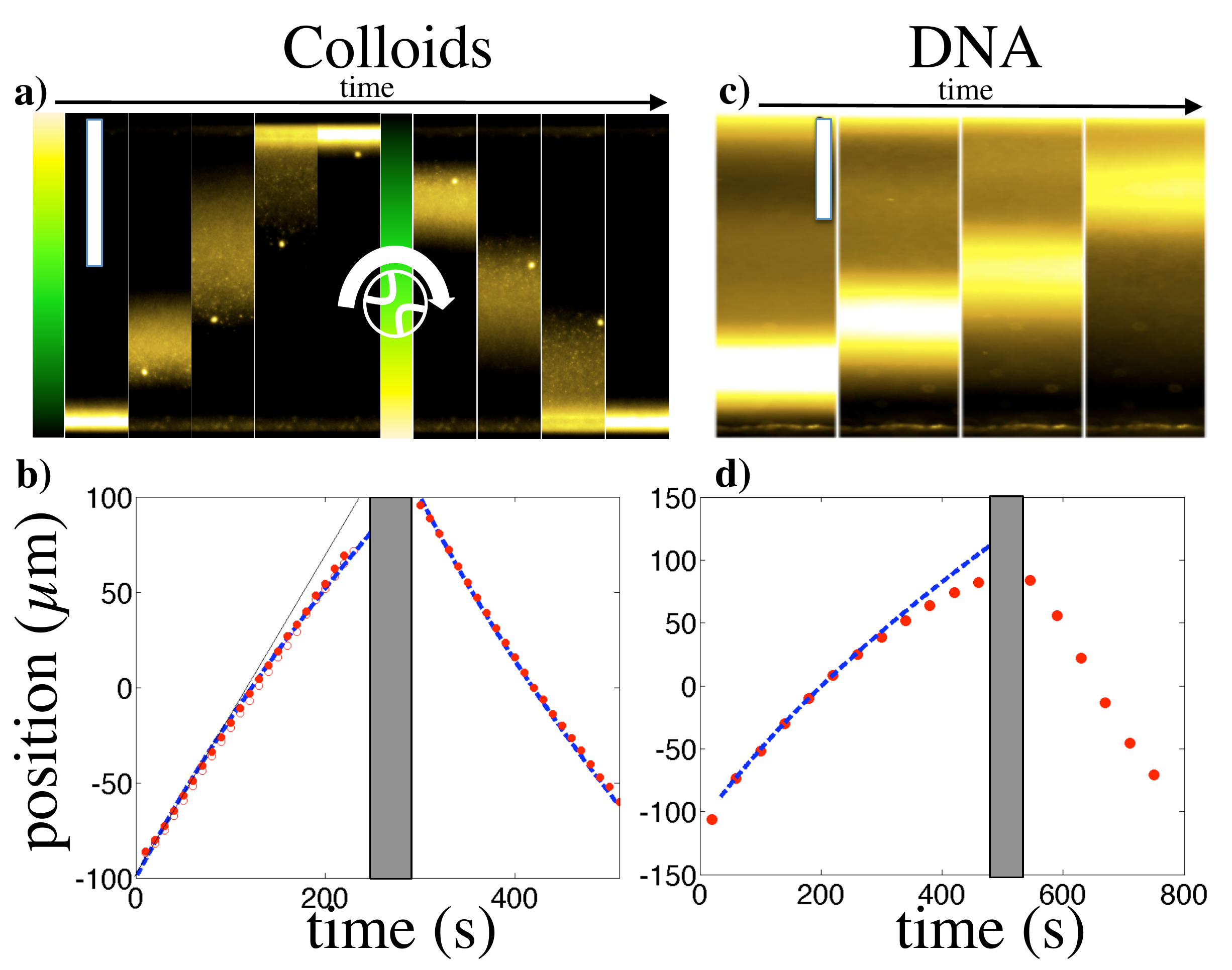}
\caption{Diffusiophoretic transport of fluorescent colloids and $\lambda$-DNA under a LiCl gradient
($\Delta C_s[LiCl]= |C_2-C_1| = 100$mM. 
$\ell=800\micron$, scale bar 100$\mu$m). {\it (a)-(c)}).  
Motion of particles under a salt gradient, sketched by the lateral bar, towards
higher salt concentrations. 
Images separated by 90s for the colloids and images at $t=$100, 150, 200, 300s for $\lambda$-DNA. {\it (b)-(d)} Time evolution of the particles population  location.
Experimental data (symbols) are 
fitted according to the theoretical description (dashed lines), with $D_{DP}$ as the only fitting parameter (see text). In (b), open symbols correspond to a subsequent migration, and fully superimposes on the previous results. The solid straight line in (b) is a guide line corresponding to constant drift velocity. Shaded regions correspond to time periods where wall effects prevent from proper fluorescence measurements.
\label{figure-2}
}
\end{figure}

Let us first explore the response of colloidal particles under stationary salt gradients.
Starting from a configuration with all particles gathered on one side of the channel, the salt gradient is switched 
 toward the opposite channel wall: 
 colloids are accordingly observed to drift toward the higher solute concentration, 
see Fig.~\ref{figure-2}-(a). 
In order to quantify this motion, we plot in Fig.~\ref{figure-2}-(b) the time dependent location of the colloidal population, 
defined as the maximum of the distribution. 
The observed drift is close to linear, as would be expected for a constant mobility $\mu=V_{DP}/\nabla c$. However a slight deviation from this linear expectation can be observed at long times. 
This result can be understood by going more into details of the diffusiophoretic transport. Indeed, in the case of electrolytes as a driving solute,
the mobility $\mu$ is expected to depend on the salt concentration $c$, scaling as $\mu(c) \sim c^{-1}$,
and the diffusio-phoretic velocity under a salt concentration gradient $\nabla c(x,t)$ can be written as \cite{Prieve87}:
\begin{equation}
V_{DP}=\DDP \nabla \log c
\label{VDP}
\end{equation}
with $\DDP$ a diffusio-phoretic mobility  \cite{anderson}. 
Physically, this dependence originates in the balance between osmotic forces and visous stresses occuring within the Debye layer at the particle surface \cite{AjdariBocquet2006,abecassis2008}. This leads
to $\DDP$ scaling typically as $\DDP \sim {k_BT/ \eta\,
\ell_B}$, with $\eta$ the water viscosity and $\ell_B$ the Bjerrum length ($\ell_B =0.7$nm in water). 

This non-linear dependence on solute concentration is at the origin of the deviation from constant drift observed experimentally.
Indeed under a linear salt profile 
$c(x)={c_0 / 2} (1 \pm 2 {x/ \ell})$, with $\ell^{-1}$ the slope of the concentration gradient ($\ell^{-1}=\nabla c/c_0$), and $x$ the distance to 
mid-channel, Eq.~(\ref{VDP}) predicts that 
the position $X_0$ of a colloid obeys  
%
${dX_0/ dt}= \pm {\DDP}/{( {\ell}/{2} \pm X_0)}$.
%
This equation can be solved analytically and
provides a very good fitting expression for experimental data,  see Fig.~\ref{figure-2}-(b),
allowing for the determination of the colloids diffusio-phoretic mobility $D_{DP}$.
In the case of colloids, the diffusiophoretic mobility was measured for three different salts, exhibiting salt specificity effects with values for the mobility in the order LiCl$>$NaCl$>$KCl, see Table I, in line with previous results \cite{abecassis2008}. 
Furthermore, we performed additional experiments with $\lambda$-DNA in place of spherical colloids: Figs.~\ref{figure-2}-(c,d) demonstrate a comparable motion of DNA molecules
resulting from 
the diffusiophoretic migration 
towards higher concentrations of salt, Table I. 

\begin{table}[t]
\begin{tabular}{|c|c|c|c|c|c|c|}
  \hline
 \multicolumn{2}{|c|}{} & \multicolumn{3}{|c|}{Colloid}& DNA \\
 \hline
 \multicolumn{2}{|c|}{Salt} & LiCl & NaCl & KCl & LiCl\\
 \hline
 \multicolumn{2}{|c|}{$D_{DP}$ ($\mu m^2/s$)} &290 $\pm$ 5 & 150 $\pm$ 10 &  70 $\pm$ 10 & 150 $\pm$ 20\\
 \hline
 \multirow{2}{2.5cm}{$(K\ell^2/k_BT)^{1/2} $} & Exp. & 22 $\pm$ 2 & 16$ \pm$ 1.5 & N.A. & 12.5 $\pm$ 3\\
 \cline{2-6}

 & Theo. &   23$\pm$ 0.5 & 16.5$\pm$1& 13$\pm$2 & 16.5 $\pm$ 1\\
 \hline
 \end{tabular}
\caption{Values for the diffusiophoretic mobility $\DDP$ extracted from Fig. \ref{figure-2} (see text).
Experimental trap strength $K$ extracted from the trapping experiments in Fig. \ref{figure-3}, and compared to the predicted value from
Eq. (\ref{gaussian}). \\}
\end{table}

\begin{figure}[tbp]
\includegraphics[width=8cm]{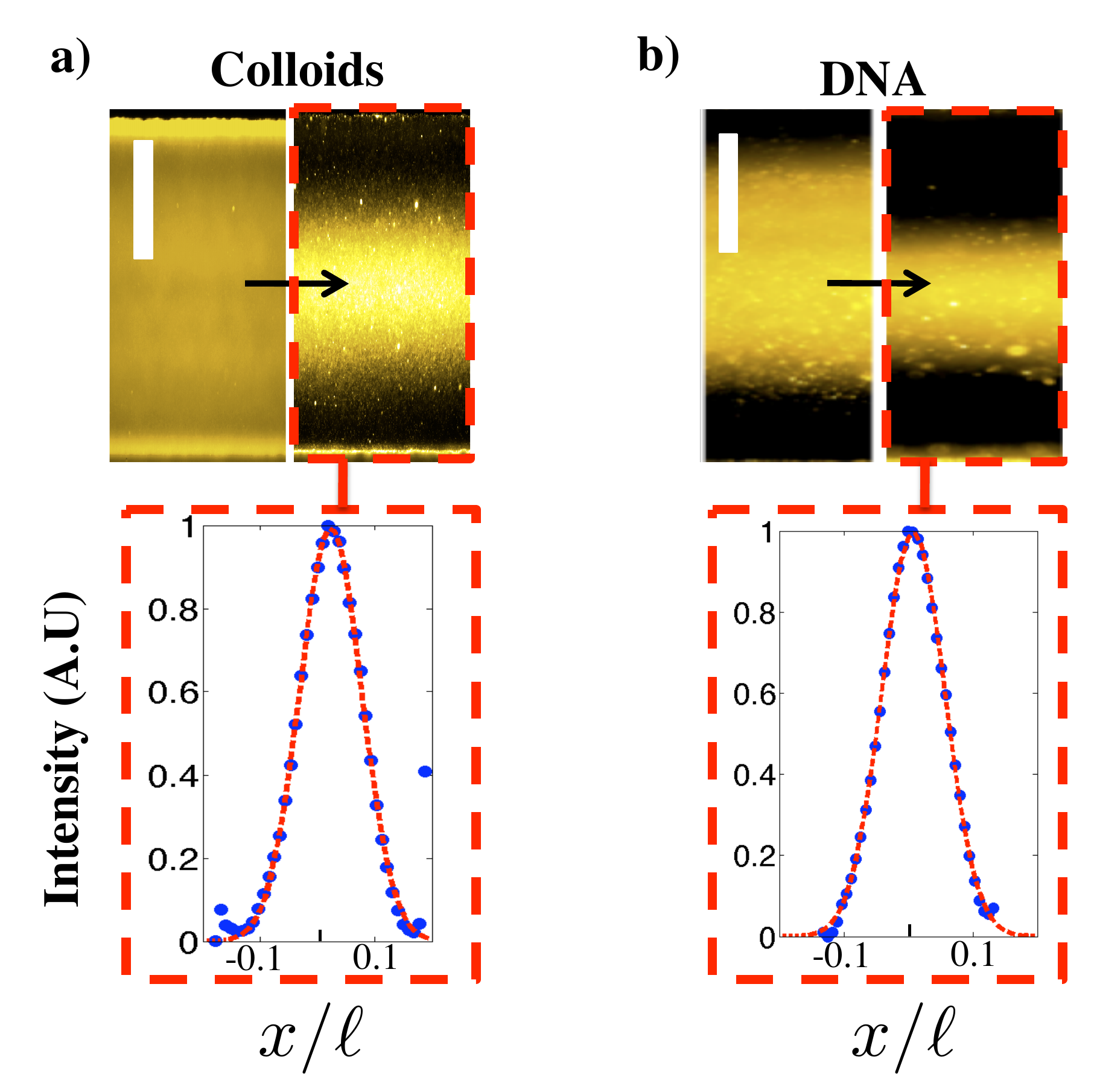}
\caption{Trapping under oscillatory gradients: {\it (Top)}: trapping of colloids, (a), and $\lambda$-DNA, (b), under salt gradient oscillations. Buffers and salt concentration are identical to  Fig.~\ref{figure-2} with $T_0=600$s (scale bar 100$\mu$m). 
({\it Bottom}):  experimental particles density profile in stationary state, as obtained from the fluorescence images (symbols), fitted to a gaussian curve (dashed line). 
\label{figure-3}
}
\end{figure}

We now turn to the effect of solute time-dependent oscillations:
while in the previous section migration  was studied under a stationary gradient, we now generate concentration gradients oscillations -- Fig.~\ref{figure-1}-(b) -- 
(with period $T_0$ in the range $T_0/2\sim 180 -300$s in the following).
Typically, experiments show that a linear solute gradient establishes in the system within a time $T_t \sim 50$s for $\ell=800\mu$m, in agreement
with numerical estimates \cite{Supp}.
The result 
is shown in Fig.~\ref{figure-3}-(a)  for colloids: starting from a homogeneous distribution, 
the particles population evolves after a few cycles 
towards a band of {\it gaussian shape} with stationary width, Fig.~\ref{figure-3}-(a) \cite{Supp}.
Note that the position of the band keeps oscillating around the center with the salt cycles, but its width remains stationary. Moreover, the process is robust: starting from inhomogeneous particles distribution, or changing the oscillation frequency 
yields the same width for the trapped band. Finally as for the colloids motility, the width of the trapped band is dependent on the salt nature: LiCl $>$ NaCl $>$ KCl in terms of trapping efficiency.
Additionally, we also performed similar experiments with $\lambda$-DNA as motile particles: Fig.~\ref{figure-3}-(b) shows that trapping is equally achieved with macromolecules, the latter gathering also in a narrow band with gaussian profile.

Going further, a physical interpretation for the segregation can be proposed: 
as we now show, trapping of the particles indeed results from the rectification of their motion under the oscillating driving field of the salt concentration. 
The particles' population obeys the Smoluchowski equation
\begin{equation}
\partial_t \rho = -{\nabla}\cdot \left( - D_c \nabla\rho_0 + \DDP \nabla[\log c] \times \rho_0 \right),
\label{Smolu}
\end{equation}
coupled to the solute diffusive dynamics on $c(x,t)$.
Neglecting transients in the salt dynamics,
its concentration profile can be written as $c(x,t)=c_0(1/2 + f(t)\, x/\ell)$,  
with $f(t)$ a time-dependent function, oscillating with the forcing periodicity, $T_0$.  
Typically, $f(t)$ can be approximated as a $\pm 1$ step-function, 
so that $\langle f(t) \rangle=0$ while $\langle f(t)^2 \rangle=1 \ne 0$, with $\langle\cdot\rangle$ the time average over $T_0$.
Accordingly the position $X_0(t)$ of the particle population is expected to oscillate with the gradients following the
description above and Eq. (\ref{VDP}). In the limit of
small excursions around the center ($X_0\ll\ell$), this reduces to  $dX_0/dt \simeq 2\DDP f(t)/\ell \left[1 - 2 f(t) (X_0/\ell) + \ldots\right]$. 

Now, in the limit of fast salt oscillations,
one expects the distribution in steady state to behave to leading order as $\rho(x,t)\simeq\bar\rho_0(x-X_0(t))$.
Inserting this guess into the Schmoluchowski equation, Eq. (\ref{Smolu}) and using the above equation for $X_0(t)$
leads to the following equation for the particle density profile $\bar\rho_0$:
\begin{equation}
\langle J\rangle= - D_c \nabla_{x} \bar\rho_0 + {4\DDP\over\ell^2}\langle f^2\rangle\times  \delta x\, \times \bar\rho_0 \simeq 0
\end{equation}
with $\delta x=x-X_0(t)$. In deriving this equation, we have furthermore averaged out over the fast salt variables.
Solving this equation in the stationary state predicts a gaussian distribution for the particles:
\begin{equation}
\bar\rho_0(\delta x) \propto e^{-{\delta x^2\over 2\sigma^2}}\,\,\,{\rm with}\,\,\, \sigma=\frac{1}{2 }\sqrt{{D_c\over \DDP}} \times \ell,
\label{gaussian}
\end{equation}
that oscillates as a whole around the channel central position. 
The model thus reproduces the experimental observations: trapping towards a {\it gaussian distribution}, with a {\it frequency independent width}.
We furthermore assessed the validity of this description by performing a full numerical resolution of the coupled particle and solute dynamics
\cite{Supp}.

Physically the origin of the focusing lies in the non-linear dependency of the diffusio-phoretic phenomenon versus the solute concentration, Eq. (\ref{VDP}).
Under a constant solute gradient, the velocity of particles is larger in regions with smaller solute concentration. Accordingly, the front particles
move slower than the back ones: iterated over the oscillations, such a process leads to the observed focusing. The balance with Brownian motion
leads to 
an harmonic ``osmotic'' trapping potential,
%
${\cal V}_{\rm trap} (x)= {1\over 2} K x^2$, 
%
with $K=\kt/\sigma^2$. The experimental values for the measured trapping strength $K$ are gathered in Table~I for various salts. These values are in good agreement with theoretical predictions, in which $\DDP$ was set to the mobility measured independently in channel-crossing experiments (Fig.~\ref{figure-2}). 
A final important remark is that the trapping potential is indeed strong: over the scale of the system $\ell$, the trapping free-energy well has a depth of $\Delta {\cal F} \approx K \ell^2 = \kt\times {\DDP\over D_c}$ up to hundreds of $\kt$ ! (and independent of $\ell$).

%
%
%
%
\begin{center}
\begin{figure}[t!]
\includegraphics[width=8cm]{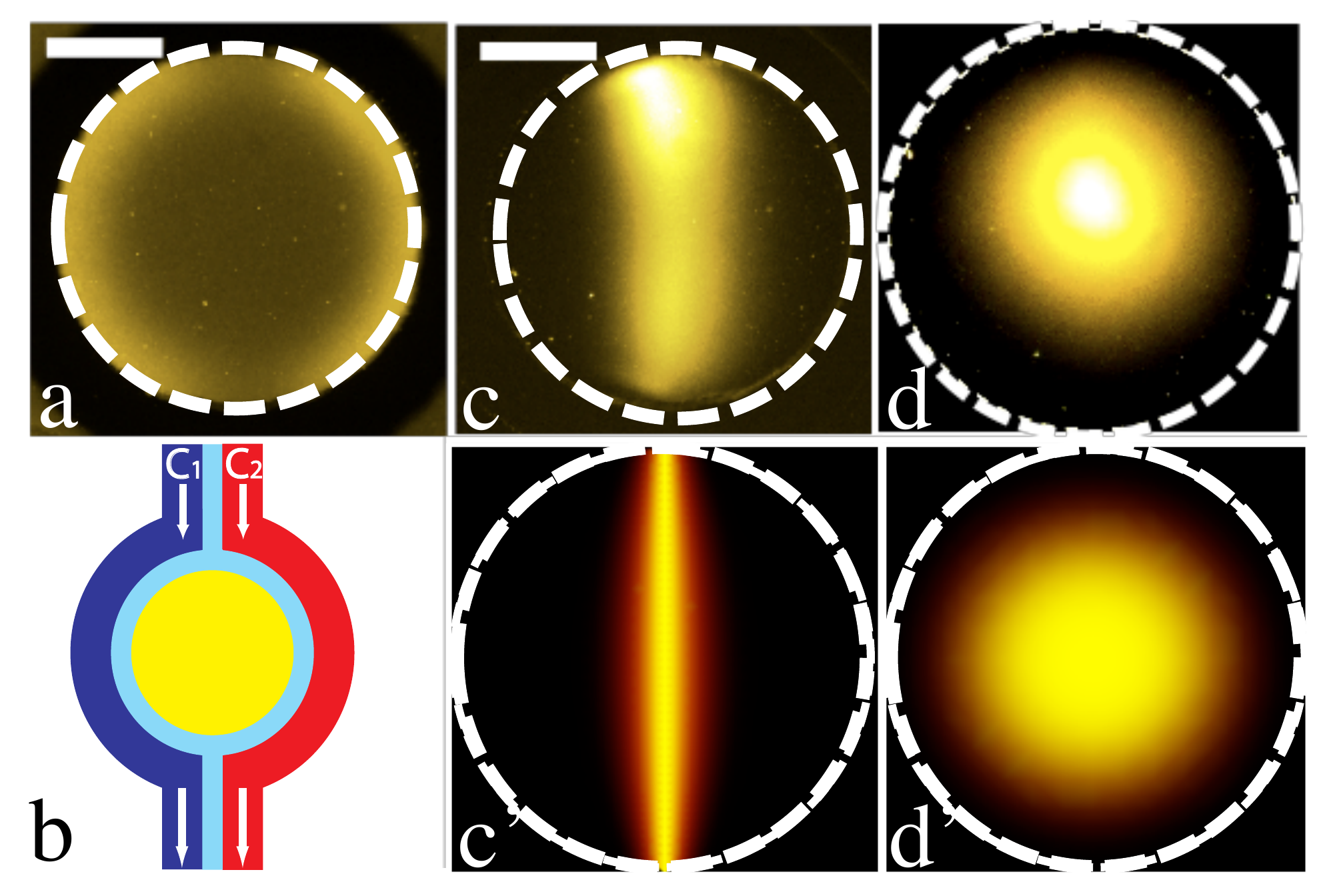}
\caption{Trapping of colloids in a circular chamber: 
\textit{(a)-(b)} Initial distribution of the fluorescent colloids 
and sketch of the experimental set-up. The central circular well is 650$\micron$ in diameter with gel walls 125 $\micron$ wide ($\ell=900\micron$; scale bar 200$\micron$). 
The salt (LiCl) concentration
oscillates in the two side channels (period $T_0=480$s) either anti-symmetrically, $C_1(t)-\langle{C}_1\rangle= -[C_2(t)-\langle{C}_2\rangle]$, or symmetrically $C_1(t)=C_2(t)$, between $c_0=100mM$ and $0$ (in addition to a TRIS buffer). 
\textit{(c)-(d)}: Stationary colloidal distribution under an \textit{antisymmetric} driving (c) and under \textit{symmetric} driving (d).
 \textit{(c')-(d')}: Theoretical predictions under corresponding experimental conditions. The predicted profiles are those obtained by avering out the oscillating salt distribution (see text). The values for $\DDP$ were taken from experiments in Fig. \ref{figure-2}, see Table I, while $\ell$ and $T_0$ take their experimental values. 
\label{figure-5}
}
\end{figure}
\end{center}
We now generalize the previous results to more complex geometries in order to demonstrate the robustness and generic character of the above scenario.
To this end, we tested the localization phenomenon in a circular, ``cell shaped'', chamber, Fig.~\ref{figure-5}-(a)-(b).
Two different drivings are explored: an {\it antisymmetric} driving, where the salt boundary concentrations, $C_1$ and $C_2$ are switched periodically between 0 and $c_0$ with antisymmetric phase; and a {\it symmetric} driving, where $C_1=C_2$ switches periodically between 0 and $c_0$.
As demonstrated in Figs.~\ref{figure-5}-(c),(d), the solute oscillations again produce a localization of the particles in the cell chamber \cite{Supp}.
Furthermore the symmetry of the pattern depends directly on the symmetry of the driving: linear --``cat eye'' shape-- for the antisymmetric driving, and circular --cell-shape-- for the symmetric driving, thus demonstrating the versatility of the trapping process. A further remark is that this segregation process is also robust w.r.t. the initial distribution of the particles.

A rationalization of the different patterns observed 
can be obtained along similar lines as above. First the solute diffusion equation is solved with the appropriate boundary conditions under periodic inversion. In the case of the antisymmetric geometry, the salt concentration profile $c(r,\theta,t)$ (with $r,\theta$ the orthoradial coordinates) is found to take the general form
$c(r,\theta,t)={c_0/ 2}\left[ 1+\delta_{\rm anti}(r,\theta) \,f(t) \right]$ with $\delta_{\rm anti}(r,\theta)={4\over \pi} \sum_{k, odd} {k^{-1}} \left({r/\ell}\right)^{k} \sin[k \theta]$, and $f(t)$ the periodic step-like function with period $T_0$.
In the case of a symmetric driving, one finds $c(r,\theta,t)={c_0/ 2}\left[ 1+\delta_{\rm sym}(r,\theta,t) \right]$ where 
$\delta_{\rm sym}(r,t)={4\over \pi} \sum_{k, odd} (1/ k) {\rm Im}\left[f_{k}(r) \exp( j\, k \omega\, t)\right]$ with $f_k(r)=I_0(r/\delta_k)/I_0(\ell/2\delta_k)$, $\delta_k=\sqrt{ {j D_s/ k\, \omega}}$ the salt diffusive length, $\omega=2\pi/T_0$ and $I_0(x)$ is the Bessel function of order 0. 
In both cases, averaging out the fast salt variables,
the trapping potential is then obtained in the form ${\cal V}_{\rm trap}(r,\theta)={1\over 2} (\DDP/ D_c)\times  \phi(r,\theta)$: 
$\phi(r,\theta)=\delta_{\rm anti}(r,\theta)^2$ for the antisymmetric case, 
while $\phi(r, \theta)\simeq \kappa_0\, r^4$ for the symmetric driving, with $\kappa_0^{-1}= 16\pi \delta_1^5\cosh(\ell/\sqrt{2}\delta_1)/\ell$.
These predicted localization patterns are exhibited in Figs.~\ref{figure-5}-(c$^\prime$),(d$^\prime$), showing a good qualitative and semi-quantitative agreement with the experimental results. 

To conclude, we have demonstrated a new mechanism leading to segregation and pattern formation of colloids and macromolecules,
originating in rectified time-dependent solute contrasts. A key ingredient of the process is the so-called {\it diffusiophoresis}, a passive transport
phenomenon leading to motion of particles under chemical potential gradients. Our results highlight the importance of solute contrast induced transport
in out-of-equibrium processes, with potential implications in soft matter, as well as in living and chemical systems, where concentration
gradients are ubiquituous.

We thank H. Ayari, F. J\"ulicher, J.-F. Joanny and P. Jop for highlighting discussions and H. Feret for technical support.
We acknowledge support from R\'egion Rh\^one-Alpes under CIBLE program.


\begin{thebibliography}{99}



\bibitem{anderson}
J.L. Anderson, 
\newblock {\em Ann. Rev. Fluid Mech}, {\bf 21}, 61-99 (1989)

\bibitem{abecassis2008}
B. Abecassis, C. Cottin-Bizonne, C. Ybert, A. Ajdari, L. Bocquet,
\newblock {\em {Nature Mat.}}, {\bf 7} {785--789}  (2008).

\bibitem{Prieve87} D. Prieve, R. Roman, 
{\it J. Chem. Soc., Faraday Trans.}, {\bf 83}, 1287-1306 (1987).

\bibitem{Prost2009} F. J\"ulicher, J. Prost,
\newblock{ \em{Eur. Phys. J. E}}, {\bf 29}, {27-36}  (2009)

\bibitem{Duhr06}
S. Duhr, S. Braun, 
\newblock {\em Proc. Nat. Acad. Sci. USA}, {\bf 103}, 19678--19682 (2006)

\bibitem{Jiang09}
H.R. Jiang, H. Wada, N. Yoshinaga, M. Sano,
\newblock {\em Phys. Rev. Lett.}, {\bf 102}, 208301 (2009).

\bibitem{Wurger09}
A. W\"urger, {\it Phys. Rev. Lett.}, {\bf 102} 078302 (2009)

\bibitem{Piazza08}
R. Piazza, {\it Soft Matter}, {\bf 4} 1740 (2008)

\bibitem{AjdariBocquet2006}
A. Ajdari, L. Bocquet, 
\newblock {\em Phys. Rev. Lett.}, {\bf 96}, 186102 (2006).


\bibitem{abecassis2009}
Abecassis B., Cottin-Bizonne C., Ybert C., Ajdari A., Bocquet L. 
\newblock {\em New J. Phys.} {\bf 11} 075022 (2009).

\bibitem{Kabalnov09} A. Kabalnov and H. Wennerstr\"om, {\it Soft Matter} {\bf 5} 4712 (2009).

\bibitem{Golestanian} 
J. Howse {\it et al.}, {\it Phys. Rev. Lett.} {\bf 99} 048102 (2007)

\bibitem{Wu06}
J. Diao,  {\it et al.} 
\newblock {\em Lab on a Chip}, {\bf 6}, 381-388 (2006).






\bibitem{Supp} See supplementary material at http://link.aps.org/
supplemental/10.1103/PhysRevLett.104.138302.

\end{thebibliography}
\end{document}